# A high-mass Planetary Nebula in a Galactic Open Cluster

V. Fragkou
Department of Physics & The Laboratory for Space Research, University of Hong
471 Kong, Pok Fu Lam Road, Hong Kong;
e-mail: vfrag@hku.hk

Q. A. Parker*
Department of Physics & The Laboratory for Space Research, University of Hong
471 Kong, Pok Fu Lam Road, Hong Kong;
e-mail: quentinp@hku.hk

A. A. Zijlstra
Jodrell Bank Centre for Astrophysics, School of Physics & Astronomy, University of Manchester,
Oxford Road, Manchester M13 9PL, UK; Department of Physics & The Laboratory for Space
Research, University of Hong Kong, Pok Fu Lam Road, Hong Kong;
e-mail: a.zijlstra@manchester.ac.uk

L. Crause
South African Astronomical Observatory,
P.O. Box 9, Observatory, 7935, South Africa
e-mail: lisa@saao.ac.za

H. Barker
Jodrell Bank Centre for Astrophysics, School of Physics & Astronomy, University of Manchester,
Oxford Road, Manchester M13 9PL, UK;
e-mail: helen.barker@postgrad.manchester.ac.uk

**ABSTRACT**

**Planetary Nebulae are the ionised ejected envelopes surrounding the remnant cores of dying stars. Theory predicts that main-sequence stars with one to about eight times the mass of our sun may eventually form planetary nebulae. Until now no example has been confirmed at the higher mass range. Here we report that planetary nebula BMP J1613-5406 is associated with Galactic star cluster NGC 6067. Stars evolving off the main sequence of this cluster have a mass around five solar masses. Confidence in the planetary nebula-cluster association comes from their tightly consistent radial velocities in a sightline with a steep velocity-distance gradient, common distances, reddening and location of the planetary nebula within the cluster boundary. This is an unprecedented example of a planetary nebular whose progenitor star mass is getting close to the theoretical lower limit of core-collapse supernova formation. It provides evidence supporting theoretical predictions that 5+ solar mass stars can form planetary nebulae. Further study should provide fresh insights into stellar and Galactic chemical evolution.**

## Introduction and Background

Stars live their lives as nuclear fusion reactors and their fate usually depends on birth mass. Massive stars burn their fuel quickly and can explode as supernovae after a few million years. The vast majority of stars have lower mass and live for many billions of years. Planetary Nebulae (PNe) may eventually form from stars of ~1-8 $M_\odot$. Such PNe progenitors represent 90% of all stars more massive than the sun. Towards the end of their lives most such stars pass through the Asymptotic Giant Branch (AGB) phase where the bulk of mass-loss occurs. A final, ejected envelope is ionized by the UV radiation field from the hot, remnant central star (CS) forming a

PN. PNe can be detected to great distances where their strong emission lines permit determination of PN expansion velocity and age, so probing the physics and timescales of stellar mass loss[1]. They can be used to estimate CS luminosity and temperature and the elemental abundances of the expelled gas. PN formation rates give the death rate of low-to-intermediate mass stars and they directly probe Galactic stellar and chemical evolution[2]. Their different shapes (morphologies) provide clues to their formation, evolution, mass-loss processes and shaping action of binary central stars[3]. As the CS fades to a White Dwarf (WD) and the nebula expands, the integrated flux, surface brightness and radius change in ways predictable via hydrodynamic theory[4]. PNe are consequently powerful astrophysical tools, providing key insights into late stage stellar and chemical evolution.

Obtaining accurate distances to Galactic PNe has proven to be very difficult in the past, except for the few with CS trigonometric parallaxes. Gaia DR3[5] promises to rectify this for many Galactic CSs but many PNe CS will remain too faint for Gaia. Another key problem is the unknown mass and abundance of their Main Sequence (MS) progenitor stars that could be linked to PN evolution and chemistry[6].

These issues are overcome for PNe within Galactic star clusters whose stars were formed coevally from the same interstellar cloud and so share the same chemical environment. Indeed, a key motivation for uncovering cluster PNe was to benchmark PNe distances. Cluster distances, ages and PN progenitor masses are well determined from cluster Colour-Magnitude Diagrams (CMDs). Initial progenitor and final CS masses can be used as independent points for the metallicity dependent[7] fundamental WD Initial-to-Final Mass Relation (IFMR)[8] that correlates WD properties to their progenitor stars, enabling tracing of nitrogen and carbon in entire Galaxies.

PN-Cluster associations are rare. Only five PNe are proven members of Galactic clusters. Four are associated with globular clusters[9] and one with an Open Cluster (OC) with a turn-off (TO) mass of 2.2 $M_\odot$[10]. We report confirmation of another cluster-PN physical association: PN BMP J1613-5406, within young OC NGC6067. This example pushes the observationally confirmed PN formation limit to much higher mass and provides an exceptional opportunity for detailed study of stellar evolutionary models. We set out below evidence that this association is real based on concordance between PN and cluster of distance, reddening, PN-cluster angular and physical location within the cluster boundary and, crucially, radial velocity. The PNe physical size, chemistry and morphology are also as expected for a high mass progenitor in such a cluster.

BMP J1613-5406 (PNG 329.8-02.1) is a low surface brightness, bipolar PN discovered from the SuperCOSMOS Hα Survey (SHS)[11] and listed in the Hong-Kong/AAO/Strasbourg Hα (HASH) Planetary Nebula Catalogue[12]. The PN is located in the northern region of Galactic open cluster NGC 6067. Its centre is at RA: 16h13m02s, Dec: -54°06'32'' (J2000) and its major and minor axes are 335 and 215 arcseconds. The PN's proximity to the cluster's center (~7 arcminutes) drew our attention to possible membership. Fig.1 gives an image montage of the Cluster and PN from the SHS on-line data. The adopted PN integrated Hα flux is $\log F_{H\alpha}$= -11.55 mW/m$^2$, the average of values measured from the Southern Hα Sky Survey Atlas (SHASSA) and the SHS data[13,14] in Table 2.

NGC 6067 is a well-studied, young OC at α= 16h13m11s and δ= -54°13'06" (J2000). It contains two classical Cepheid variable stars[15] that provide independent distance estimates and other insights into mass loss processes in later stellar evolution. Another Planetary Nebula, HeFa1, resides ~12 arcmin South-West of the cluster's centre. This is a proven chance alignment[16].

Taking the error weighted average of published cluster parameters (see Table.1) we find an age of 90 ± 20 million years, a distance of 1.88 ± 0.10 kpc, a reddening of E(B-V)= 0.35 ± 0.03, a tidal radius of 12.3 arcminutes and a heliocentric radial velocity of -39.79 ± 0.57 km/s. These are in excellent agreement with findings of the most recent cluster study[17]. The latest cluster abundance studies[17] predict a cluster metallicity of [Fe/H] = 0.19 ± 0.05 (Z=0.024 assuming a solar abundance of $Z_\odot$=0.0152). The cluster MS TO point is estimated at a spectral type of B6[17,18]. Both the TO spectral type and estimated cluster age indicate a MS TO of ~5 $M_\odot$[19]. Theoretical cluster isochrones[20-22] for the adopted cluster parameters and accounting for the time since the star left the MS and passed through the AGB phase, predict a PN progenitor mass of $5.58^{+0.62}_{-0.43}$ $M_\odot$. Such a high mass for a PN progenitor is unprecedented and provides a key new datum at the poorly studied IFMR high-mass end.

The cluster's estimated distance is within Gaia DR2's reach[5] which provides accurate stellar distances and radial velocities for cluster members. From the Gaia DR2 parallaxes of 43 cluster stars the error weighted mean cluster distance is 1.94 ± 0.07 kpc with σ= 0.37 (one suggested cluster member is excluded on proper motion grounds). Only 12 have Gaia radial velocities. There is one outlier (likely contaminant) so the 11 remaining stars yield an error weighted average cluster radial velocity of -39.21 ± 0.15 km/s with σ= 1.01 km/s, in good agreement with the best previous literature radial velocity estimates.

1.       Observational evidence supporting Planetary Nebula-Cluster membership

Confirming a PN-cluster association requires agreement of multiple PN and cluster parameters such as position proximity within the cluster boundary, reddening, distance and crucially, radial velocity. OC velocity dispersions are typically ~1 km/s[23] so agreement here is a particularly tight constraint. NGC 6067's sightline shows a steep velocity-distance gradient (-16 ± 3 km/s/kpc)[15] making a chance radial velocity coincidence unlikely. BMP J1613-5406 lies ~7 arcminutes from the cluster center, well within its projected tidal radius of ~14 arcmin and radial extent[15]. From our six PN spectra from the High Resolution Spectrograph (HRS) on the SALT 10-m telescope and using the [NII] 6548Å, Hα and [NII] 6584Å lines, we measure a heliocentric corrected radial velocity of -39.93 ± 1.44 km/s. This precisely agrees, within the tight errors, with the cluster radial velocity (radial velocity difference, dRv<1 km/s) obtained from previous cluster studies and Gaia data.

Our ESO VLT XSHOOTER spectra of the CS and PN have the Hα and Hβ emission lines in different arms of the spectrograph so a reddening calculation from the Hα/Hβ Balmer decrement is not considered reliable even when flux calibrated. Instead PN reddening was estimated as E(B-V)= 0.38 ± 1.1 from the Hγ/ Hβ Balmer decrement as these lines fall in the same arm. The large PN reddening uncertainty means PN cluster membership credentials must depend on other, more tightly constrained criteria (e.g. radial velocity, common distance and independent CS reddening).

For calculating an independent PN statistical distance we use the robust $H_\alpha$ surface brightness-radius relation[24] and the mean Hα integrated flux of $logF_{H\alpha}$= -11.55 mW/m$^2$, found from the two published measures[13,14], corrected for the adopted PN reddening. This yields a PN distance of

$1.71^{+0.29}_{-0.24}$ kpc, in good agreement with the mean cluster distances from both Gaia parallaxes and previous cluster studies.

Agreement of the required parameters (Table 2) and projected physical proximity to the cluster centre (~1.7pc) suggests that PN BMP J1613-5406 is a physical member of NGC 6067. The likely bipolar morphology, given the rectangular shape of the clearly evolved PN, and its plausible Type-I chemistry (see section 3) are characteristics thought to derive from relatively high mass progenitors, as expected if hosted by NGC 6067.

## 2. Planetary Nebula properties

The nebular emission line fluxes and flux ratios were measured using the splot/IRAF task from our ESO VLT XSHOOTER spectra, SALT HRS spectra and our integrated AAT SPIRAL IFU spectra (see Fig. 2 and Table 3). The extinction was measured from our XSHOOTER original F(λ) data and then used for obtaining the I(λ) extinction corrected line fluxes[25]. The [NII]/Hα and [SII]/Hα ratios from the XSHOOTER and SALT data are lower limits as the sky could not be subtracted and spectra are contaminated by Hα sky emission. The low electron density $N_e$ from the XSHOOTER and SPIRAL data was calculated using the NEAT code[26] and is as expected for an old nebula.

Adopting the mean cluster distance of 1.91 kpc from Gaia and literature values and the mean PN angular diameter of 275 arcseconds, we find a physical diameter of 2.54 pc, clearly indicating its evolved nature. The PN expansion velocity of 39 km/s, typical for a PN[27], was measured from our SALT HRS data using the mean HWHM of the Hα line. The prominent [NII] 6584Å line is split in the SALT central region spectra and gives an expansion velocity of 42 km/s. Adopting a mean expansion value of 40.5 km/s and considering the estimated PN physical radius yields a nebular kinematic age of $t_{kin}$ ~30,600 years. This is also typical for an old, evolved PN[28] from an intermediate to high mass progenitor. The PN ionized mass was calculated as[29]:

$$M_{ion} = 4.03 \times 10^{-4} \epsilon^{\frac{1}{2}} d^{\frac{5}{2}} F(H_\beta)^{1/2} \Theta^{3/2} M_\odot \qquad (1)$$

where d is the distance in kpc, F(Hβ) the Hβ line flux in units of $10^{-11}$ erg cm$^{-2}$ s$^{-1}$ and Θ the average PN radius in arcseconds. Converting the derived mean integrated F(Hα)[13,14] to F(Hβ) flux using a standard line ratio of 2.85 and assuming a filling factor of $\epsilon$=0.3[29] we estimate an $M_{ion}$ ~ 0.56 $M_\odot$. Taking the adopted values for the nebular diameter, filling factor, and mass, the density of the nebula is estimated as $N_e$ ~9 cm$^{-3}$. This is consistent with the observed [S II] ratio, which is at its low-density limit.

Weak detection of the He II 4686Å line in both the SPIRAL and XSHOOTER nebular spectra indicates a high excitation PN so the CS must have a temperature of at least 50,000K. From our XSHOOTER emission line ratios we measure the excitation class parameter as $E_{Ex*}$= 5.5[4] and $Ex_\rho$= 4.3[30]. These values are lower limits as the XSHOOTER spectra are contaminated by sky lines. From the sky subtracted SPIRAL spectrum the $E_{Ex*}$ excitation class is estimated around 7.3, though a calculation of a reliable $Ex_\rho$ is not possible at it relies on emission lines present in separate spectrograph arms and our SPIRAL data are not flux calibrated. An $Ex_\rho$ > 4.3 predicts a CS temperature of at least 81,000 K[30]. Using the crossover CS temperature estimation method[31] that assumes the nebula is optically thick, we calculate a $T_{cross}$ of at least 125,000 K from the XSHOOTER spectra and 190,000 K from the SPIRAL data. The estimated CS magnitudes expected from these values[31] are $V_{cross}$ >20.3 and 21.4 respectively. From the estimated CS temperature and magnitude values and the established mean cluster-PN reddening and distance we calculate a CS luminosity of log(L/L$_\odot$) ~ 1.66-2.12. By plotting the derived kinematic age and estimated absolute CS magnitudes, correcting the visual magnitudes for the adopted distance and reddening, along with evolutionary tracks[32], and then interpolating the mass values measured from the closest plotted tracks, we expect a CS mass of ~ 0.94 ± 0.11 M$_\odot$[32]. The CS mass was estimated from the mean CS magnitude derived from the crossover method applied to both our SPIRAL and XSHOOTER data and its error reflects their spread. Our estimated PN and CS properties are summarized in Table 2.

Taking corner-to-corner diagonals across the rectangular H$_\alpha$ image of the PNe provides a decent geometric centroid position where they overlap. A very faint blue star is located within a couple of arcseconds of this position and was reported as the likely CS upon PN discovery[33] (see Fig.3). We find RA:16h13m02.1s and DEC:-54°06'32.3" (J2000) for the candidate CS and used the VPHAS+[34] survey to determine its photometric magnitudes. The u and g fluxes of all stars in the nebular field were measured with the IRAF/daophot package[35] revealing that this is the only blue star within the PN confines of 20 square arcminutes. This stars' VPHAS+ g magnitude was transformed to the V magnitude system by cross-correlating the g magnitudes of stars in the nebular field with their APASS V magnitudes. From the resulting tight linear relation, the star's V mag is ~20.2, in very good agreement with our CS V magnitude prediction from our XSHOOTER nebula data. Taken with the CS location at the PN's geometric centre, this gives high confidence that the true CS has been identified. Fitting our measured VPHAS+ u-g stellar colour to published reddening lines[34] (assuming a high temperature blackbody spectrum) we estimate a CS reddening of E(B-V)=0.28 ± 0.04[36]. This independent VPHAS+ photometric reddening estimate for this CS candidate is much more consistent, within the uncertainties, with that for the cluster and when compared to the large uncertainties for the PN reddening estimate. Our XSHOOTER spectrum also includes the candidate CS. Based on the equivalent widths of the CS He I 4471Å and He II 4542Å absorption lines the spectral type is estimated as O6, with an error of 1 subclass[37]. The stellar spectrum has a low S/N ~4 and modest wavelength resolution and does not permit measurement of a reliable radial velocity.

## 3. Discussion

Cluster membership is solid in terms of concordant radial velocities to dRv<1 km/s, (especially given the steep sightline velocity gradient), distance, CS and PN reddening and location within the cluster boundary. The PN properties are also as expected for location within such a cluster. The implications of this PN-OC association for stellar evolution are now considered.

The high PN [NII]/Hα ratio and estimated N abundance of log (N/H) = -3.02 from our XSHOOTER data (from the NEAT code[26] and applying appropriate ionization correction factors), indicate the nebula is likely of Type-I chemistry[38] with enhanced Nitrogen. Type-I PNe are usually bipolar and considered to emerge from higher mass progenitors[38] as is the case here. Their chemical abundances yield enhancements that arose during the AGB stage and imply, that apart from hot-bottom burning, intermediate mass stars suffer additional nucleosynthetic and mixing processes. Precise chemical abundances can be derived to clarify this issue. Furthermore, it will enable analysis of a post-AGB ejected envelope from what was a high mass progenitor star for the first time. Our estimated N/O ~0.44 abundance ratio also agrees with theoretical predictions of AGB chemical yields for a MS star of ~5.5 $M_\odot$[6]. Better data will enable empirical testing of the latest nucleosynthetic predictions[6,39] in heavyweight intermediate-mass stars for a range of elements.

Bipolar PNe may have thick tori but optically thin lobes[40] while Type-I PNe are typically optically thick[31], as supported by the strong N and detection of weak He emission in our PN spectra. We consider the calculated $T_{cross}$ a good approximation of the CS effective temperature. An excellent blue CS candidate is found at the PN's geometric centre with a V magnitude estimate as expected for the true CS and with a reddening that is also consistent with that for the cluster. CS confirmation and study of its composition will help delineate the boundary between C-O and O-Ne WD formation and inform the WD core-collapse supernova formation boundary, currently poorly constrained. Additional CS photometric observations will permit direct measurement of its effective temperature, luminosity and mass, adding a valuable datum at the sparsely populated intermediate-to-high mass end of the WD IFMR[8]. This recent study[8] used a self consistent sample of 60 cluster WDs which show the IFMR slope at initial masses around 3-3.65 $M_\odot$ and predict that total stellar mass loss does not change for a moderate range of metallicities (-0.15 < [Fe/H] < + 0.15)[8]. Using our final WD mass estimation of ~1 $M_\odot$ and assuming single star evolution we overplot this new point on the latest WD IFMR estimates[8] shown in Fig. 4 that also includes the only other point from a confirmed PN-OC association[41]. Our new point agrees, within the errors, with the new published IFMR trend[8] but further photometric studies are needed to precisely measure the CS final mass.

In summary, the PN's cluster membership provides tight constraints on the lower mass limit for the progenitor mass of core-collapse supernovae and also for the intermediate to high mass end of the WD IFMR. It also provides an empirical benchmark for evaluating nucleosynthetic predictions for intermediate-mass stars. Our results confirm theoretical predications that 5+ $M_\odot$

MS stars transition through the PNe phase and are nitrogen rich. PN BMPJ1613-5406 and its cluster NGC6067 can provide important insights from stellar to galaxy (chemical) evolution.

**Methods**

Various observations were obtained to evaluate PN cluster membership. First 2 x 600 sec exposures were obtained on the PN's Northern region (RA: 16h13m09.62s, Dec: -54°04'31'') on February 2$^{nd}$ 2011 with the SPIRAL Integral Field Unit (IFU) on the AAT 3.9 m telescope. A 600 sec offset sky exposure was also obtained. The SPIRAL field of view is 22.4 x 11.2 arcseconds. The 580V and 1700I gratings give resolving powers of R=1200 and R=7000 for the blue and red spectrograph arms. Data were reduced with the 2dfdr pipeline[55]. The two PN frames were combined and sky subtracted. No flux calibration was undertaken as no standard star was observed. These data provided integrated, 1-D, sky-subtracted, PN spectra across the optical range. They gave a cluster compatible radial velocity estimate to within the errors of -43 +/- 6 kms$^{-1}$. The PN spectra enabled Type-I chemistry to be established via observed lines and line ratios. It was confirmed as a high excitation PN via detection of He II 4684Å emission line. Our SPIRAL resolution was much lower than HRS on SALT, so its radial velocity estimate will not be as precise. We also obtained exposures at the centre of the PN (RA: 16h13m01.9s, Dec: -54°06'33.0'') with the XSHOOTER spectrograph on the ESO/VLT 8.2 m (program 287.D-5064) on May 17$^{th}$ 2014. We took six 650 second exposures with the blue and green (visual) arms and eight 480 second exposures with the infrared arm. The resolving powers of the blue, visual and infrared arms are 2000, 3300 and 5400 respectively. The data were processed using the XSHOOTER pipeline[56]. No sky frames were obtained. The PN spectrum provided an extinction estimate of E(B-V)= 0.38 ± 1.1, determined from the observed Hγ/Hβ Balmer decrement after flux calibration via the accompanying standard star observations. The spectra also provided improved abundance estimates that support Type-I PN chemistry. The XSHOOTER resolution is too poor to provide high precision radial velocity estimates for the PN and CS required for cluster membership compatibility. The S/N of the CS candidate was poor due to its extreme faintness (V~20.2) preventing more than basic spectral type assignment.

For precise radial velocity measures for PN BMP J1613-5406 we used the High Resolution Spectrograph (HRS) on the Southern African Large Telescope (SALT) under program 2018-1-DDT-007 (PI: Parker) on August 9th 2018. We used the low-resolution mode with R=16000 that gives <1km/s velocity precision. Six object spectra from 3 exposures were obtained where each has a nominal object and sky fibre. Due to the PN's extended nature each set of two fibres were of the PN. The object fibre positions were at the North (RA: 16h13m06.95s, Dec: -54°05'45.78", exposure time: 1500 sec), South (RA: 16h13m01.62s, Dec: -54°:09':49.83'', exposure time: 1500 sec) and central (RA: 16h12m59.63s, Dec: -54°07'13.08'', exposure time: 1000 sec) parts of the PN. The corresponding "nominal sky fibres" were 62.5 arcsec East for the first two exposures and 55 arcsec East for the third. Data were processed using the HRS PySALT/PyHRS pipeline[57]. Sky subtraction was not done as sky fibres are effectively additional nebula exposures. No standard stars were observed so the data are not flux calibrated.

**Data Availability Statement**

Our XSHOOTER data can be accessed from the ESO Science Archive Spectral Data Products (http://archive.eso.org/wdb/wdb/adp/phase3_spectral/form). The data that support the plots within this paper and other findings of this study are available from the corresponding author upon reasonable request.

SHS data can be accessed here: http://www-wfau.roe.ac.uk/sss/halpha/

VPHAS+ data can in principle be found here: http://www.vphasplus.org/

**Main References**

**Acknowledgments**

Part of this work is from data obtained from the ESO Science Archive Facility under request number 336270. Some reported observations were obtained with the Southern African Large Telescope (SALT). QAP and VF thank HKU for travel support for the SALT observations and the Hong Kong Research Grants Council for GRF research support under grants 17326116 and

17300417. VF thanks HKU for her PhD Scholarship. We thank Dr. David Frew for input to the XSHOOTER proposal and for early project work. We have been unable to get a reply from him to agree to the co-authorship that he deserves. We thank the anonymous referees who helped to considerably improve this paper.

**Author Contributions**

V.F.: Undertook the data reduction and analysis for the PN and cluster and led the paper writing

Q.A.P: co-discovered the PN, identified it as a possible cluster member, obtained much of the follow-up data on a variety of telescopes including SALT and co-wrote the paper

A.Z.: Provided scientific input and checked the paper

L.C.: Helped facilitate SALT observations via director's discretionary time

H.B.: Helped with the VPHAS+ photometry

**Competing Interests:** None

**Figures**

*Fig.1 30 x 30 arcminute images of NGC6067 & BMP1613-5406. North-East is top left. Left: BRHa tri-colour RGB image (extracted from the online UK Schmidt Telescope SuperCOSMOS Ha Survey Ha, short-Red (SR) and broad-band 'B' images); Right: continuum-subtracted (Ha-SR) image. The PN BMP1613-5406 is identified between the red lines in the continuum-subtracted image. The compact PN HeFa1 previously suggested as a possible cluster member can be seen arrowed at the bottom right of the image about 12 arcminutes from the cluster centre.*

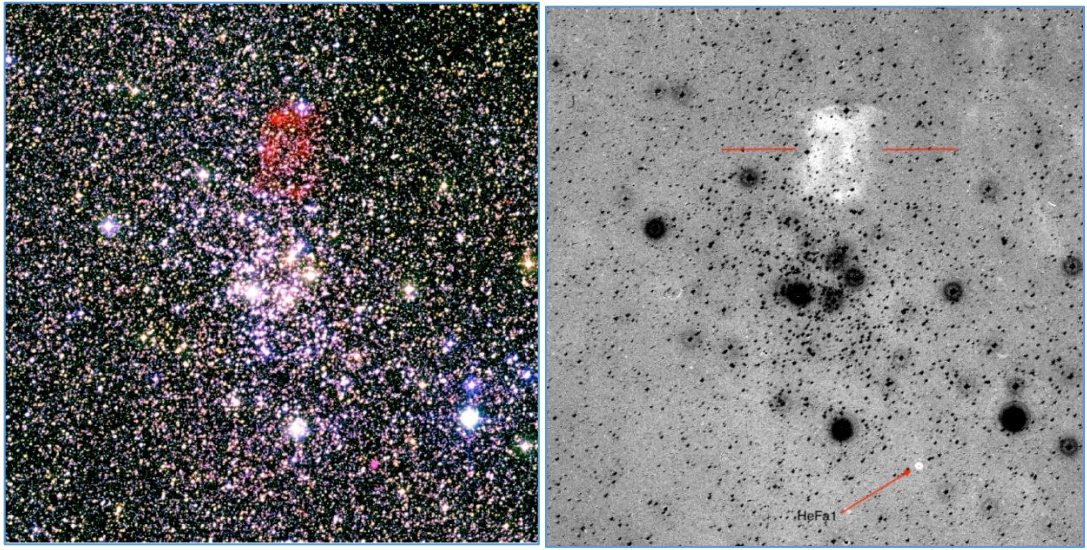

*Fig. 2 Our summed, red 3.9m AAT SPIRAL IFU 1-D spectrum with the key PN emission lines identified.*

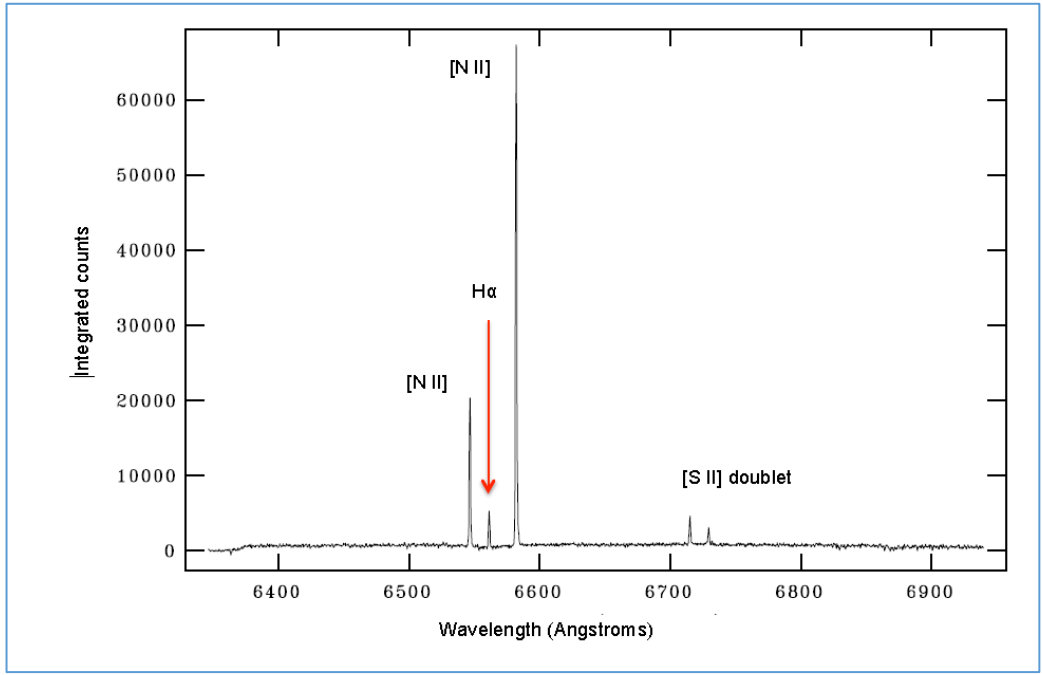

*Fig. 3. A VPHAS+ combined u g r multi-band "RGB" colour image centred on the planetary nebula central star (CS) candidate. The image is 55 x 55 arcseconds in size and the CS is obvious as the sole blue star in the middle of field located at RA:16h13m02.1s and DEC:-54º06'32.3" (J2000).*

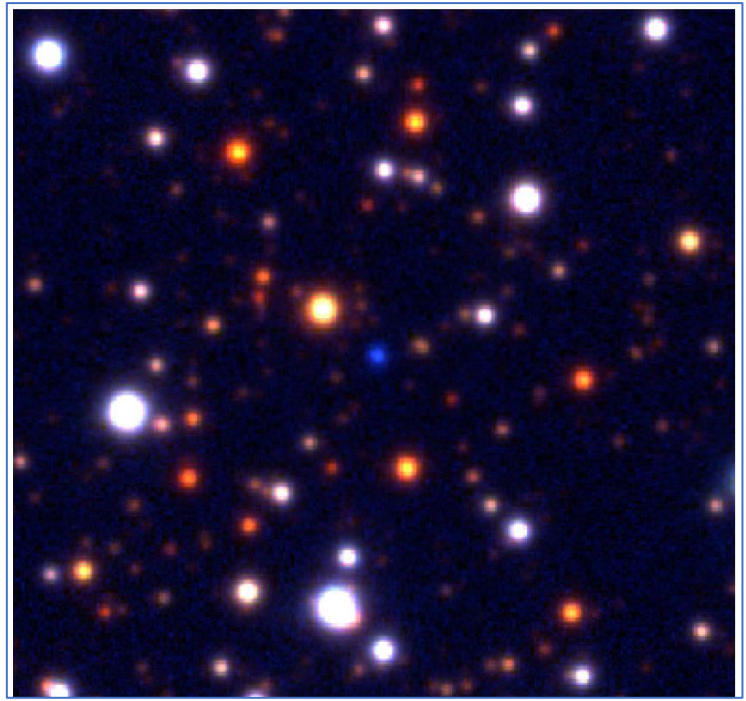

*Fig 4. A current plot from cluster WDs for the latest IFMR estimates[8] together with our estimated point for BMP1613-5406 plotted as a red circle. The only other point from a known OC PN is plotted as a yellow circle[41]. The errors attached to our point reflect the errors in the adopted cluster parameters and the spread of the estimated CS magnitudes.*

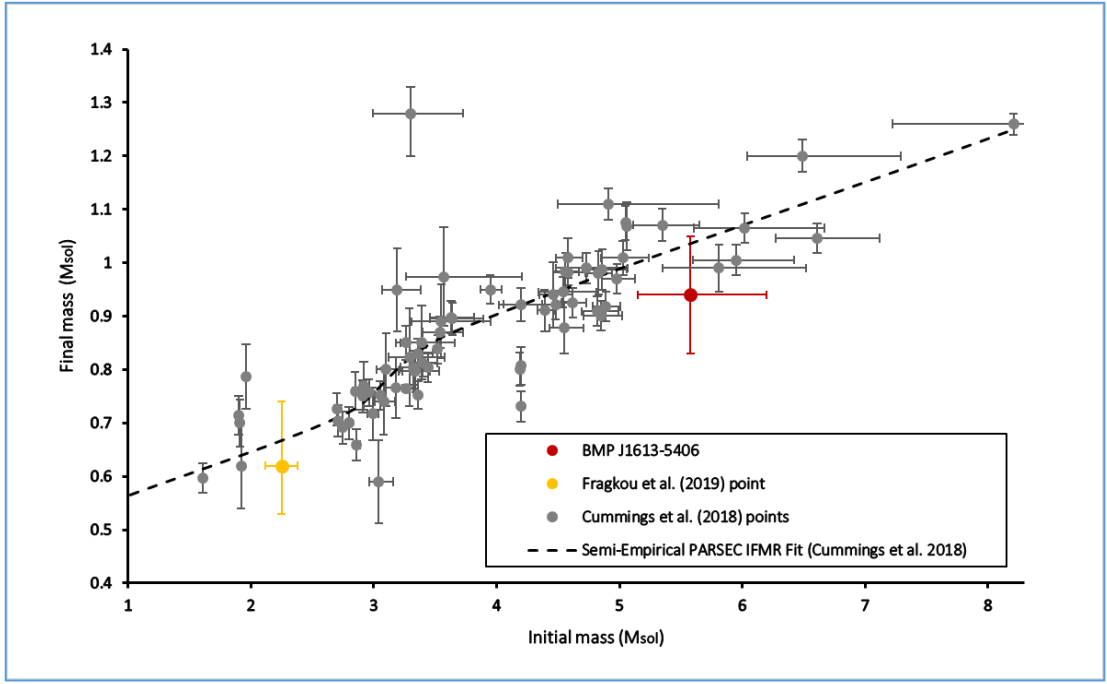

**Tables**

Table 1. Summary of published physical characteristics of NGC 6067. The presented errors are literature error values from the corresponding papers.

| Radial velocity (km/s) | Distance (kpc) | Age (Millions of years) | E(B-V) | Tidal radius (arcmin) | Reference |
|---|---|---|---|---|---|
| -39.5± 0.9 | 1.78± 0.12 | 90± 20 | 0.35± 0.04 | $14.8^{+6.8}_{-3.2}$ | [17] |
|  | 6.62± 0.16 | 100± 20 |  |  | [42] |
| -37.3± 3.02 | 1.79 | 93 | 0.412 | 14.1 | [43] |
| -39.4± 0.2 | 1.75± 0.1 | $78^{+32}_{-23}$ | 0.39[a] |  | [15] |
|  | 1.7 |  |  |  | [44] |
| -39.4± 1 | 1.6 |  | 0.34± 0.03 |  | [45] |
|  | 1.42 | 102 | 0.4 | 12.3±1.3 | [46] |
|  | 1.61± 0.06 |  | 0.34± 0.03 |  | [47] |
| -39.9± 0.16 |  |  |  |  | [48] |
| -41± 3 | 1.66± 0.08 |  | 0.35 |  | [49] |
|  | 1.6 |  |  |  | [50] |
|  | 1.62± 0.07 |  |  |  | [51] |
|  | 1.89± 0.5 | Pleiades age | 0.35 |  | [52] |
|  |  | 77.6 |  |  | [18] |
| -39.8± 0.8 | 2.1± 0.3 | Pleiades age | 0.33± 0.07 | 7.5-16 | [53] |
|  | 0.94± 0.5 |  |  | 8 | [54] |

[a] assuming E(J-H)~ 0.33 × E(B-V).

Table 2. PN-Cluster comparison of estimated physical properties.

| | BMP J1613-5406 | NGC6067 (av. previous studies) |
|---|---|---|
| Position RA (J2000) | 16h13m02s[12] | 16h13m11s[17] |
| Position Dec (J2000) | -54°06'32"[12] | -54°13'06"[17] |
| Distance (kpc) | $1.71^{+0.29}_{-0.24}$ | 1.88 ± 0.1 (1.94 ± 0.07 Gaia) |
| Radial velocity (km/s) | -39.93 ± 1.44 | -39.79 ± 0.57 (-39.21 ± 0.15 Gaia) |
| Reddening E(B-V) | 0.38 ± 1.1 | 0.35 ± 0.03 |
| Major/Minor diameter | 335 x 215 arcsec[12] | Tidal radius $14.8^{+6.8}_{-3.2}$ arcmin[17] |
| Physical radius (pc) | 1.27 | 8.1 |
| Morphology | Bipolar[33] | Open Cluster[17] |
| Chemistry | Possibly Type I | [Fe/H] = 0.19 ± 0.05[17] |
| Ionized PN mass $M_{ion}$ | 0.56 $M_\odot$ | |
| Expansion velocity | 40.5 km/s | 1km/s cluster vel dispersion[23] |
| Kinematic age $t_{tik}$ | 30600 yrs | Cluster age: 90 ± 20 Ma |
| Excitation class $E_{Ex*}$ | ~ 7.3 | |
| Excitation class $Ex_\rho$ | > 4.3 | |
| CS V mag | 20.3-21.4 | |
| CS temperature $T_{cross}$ / $T_{Exp}$ | 125-190 kK / >81kK | |
| CS Luminosity Log(L/$L_\odot$) | 1.66-2.12 | |
| CS mass | ~0.94 ± 0.11 $M_\odot$ | |
| $H_\alpha$ integrated flux from SHASSA data[13] | -11.60 ± 0.09 mW/m² | |
| $H_\alpha$ integrated flux from SHS[14] | -11.50 ± 0.12 mW/m² | |

Table 3a. Emission line fluxes and flux ratios for BMP J1613-5406 measured from our VLT XSHOOTER and AAT SPIRAL spectral data.

| Instrument | XSHOOTER | | SPIRAL |
|---|---|---|---|
| RA (h m s) | 16:13:02 | | 16:13:10 |
| Dec (° ′ ″) | -54:06:33 | | -54:04:31 |
| Line ID | F(λ) | I(λ) | I(λ) |
| [OII] 3727 | 86 | 117 | |
| [OII] 3729 | 111 | 152 | |
| H9 3835 | 11 | 14 | |
| [NeIII] 3868 | 27 | 35 | |
| Hδ 4101 | 20 | 24 | |
| Hγ 4340 | 38 | 45 | |
| HeII 4684 | 20 | 21 | 54 |
| Hβ 4861 | 100 | 100 | 100 |
| [OIII] 4959 | 41 | 39 | 32 |
| [OIII] 5007 | 90 | 86 | 257 |
| [NII] 6548 | 146 | 103 | 934 |
| Hα 6563 | 339 | 239 | 272 |
| [NII] 6584 | 370 | 260 | 3165 |
| [SII] 6716 | 120 | 83 | 184 |
| [SII] 6731 | 86 | 60 | 116 |
| [ArV] 7006 | 16 | 11 | |
| [ArIII] 7136 | 20 | 13 | |
| P9 9230 | 59 | 30 | |
| [SIII] 9530 | 94 | 47 | |
| [NII]/H$_\alpha$ | 1.52 ± 0.26 | | 15.1 ± 3.2 |
| [OIII] 5007/4959 | 2.18 ± 0.67 | | 8.0 ± 3.9 |
| [SII] 6716/6731 | 1.39 ± 0.15 | | 1.6 ± 0.3 |
| [SII]/H$_\alpha$ | 0.60 ± 0.10 | | 1.1 ± 0.2 |
| cHβ | 0.49 ± 1.38 | | |
| N$_e$ (cm$^{-3}$) | $59.4^{+114.3}_{-58.4}$ | | $1^{+50.8}_{-0.0}$ |

Table 3b. Emission line intensities and line ratios for BMP J1613-5406 measured from our SALT HRS spectral data for each of our 6 fibre spectra across 3 regions of the PN as indicated.

| Instrument | SALT | | | | | |
|---|---|---|---|---|---|---|
| | North region | | South region | | Central region | |
| | Fibre 1 | Fibre 2 | Fibre 1 | Fibre 2 | Fibre 1 | Fibre 2 |
| RA (h m s) | 16:13:08 | 62.5 ′ E | 16:13:03 | 62.5 ′ E | 16:12:59 | 55′ E |
| Dec (° ′ ″) | -54:04:40 | | -54:08:39 | | -54:06:08 | |
| Line ID | I(λ) | I(λ) | I(λ) | I(λ) | I (λ) | I (λ) |
| [NII] 6548 | 98 | 27 | 56 | 29 | 75 | 41 |
| Hα 6563 | 100 | 100 | 100 | 100 | 100 | 100 |
| [NII] 6584 | 315 | 92 | 182 | 108 | 312 | 165 |
| [NII]/ H$_\alpha$ | 4.1 | 1.2 | 2.4 | 1.4 | 3.9 | 2.1 |